# Towards an Assessment-oriented Model for External Information System Quality Characterization

Abir ELMIR[1], Badr ELMIR[2] and Bouchaib BOUNABAT[3]

Al-Qualsadi Research & Development Team
Ecole Nationale Supérieure d'Informatique et d'Analyses des Systèmes ENSIAS,
Université Mohammed V – Souissi, Rabat Maroc

**Abstract**

Information System Quality (ISQ) management discipline requires a set of assessment mechanisms to evaluate external quality characteristics that are influenced by the environmental parameters and impacted by the ecosystem factors. The present paper suggests a new assessment oriented model that takes into consideration all facets of each external quality feature. The proposed model, named RatQual, gives a hierarchical categorization for quality. RatQual is designed to quantify dependent-environment qualities by considering internal, external and in use aspects. This model is supported by a tool that automates the assessment process. This tool gives assistance in quality evolution planning and serves for periodical monitoring operations used to enhance and improve information system quality.

## 1. Introduction

To support inter organizational collaboration, there is an increasing trend for several information systems to span boundaries between organizations. While information systems quality (ISQ) enhancement is a mean to increase performance and to achieve strategic goals, many investments are done in order to improve quality levels. Studying these systems implies a specific focus on external information system quality (EISQ) characteristics influenced by the environmental parameters and impacted by the ecosystem factors.

Therefore, the domain of ISQ management is a subject to numerous engineering and research works. Many efforts are done in order to propose approaches to ensure and control quality. One of the active branches deals with the characterization and the assessment techniques. Indeed, many models are proposed to describe quality attributes and their relationships. Despite the richness and benefits of these models, they present several disadvantages regarding the consideration of all operational aspects in the characterization and assessment processes.

This work aims to propose a new assessment oriented model designed to characterize EISQ. This model clears the way to have adequate mechanisms in order to assess EISQ characteristics. Such mechanisms take into account internal, external and in use aspects.

This paper is organized as follows. The next section reminds the principal axis of EISQ management discipline. It introduces also ISQ assessment approaches. Section 3 is in relation with the main contribution of this work. It describes the new model proposed to characterize and assess EISQ. This section proposes a categorization for EISQ characteristics based on quality requirement management perspective. It notes the importance to take into consideration internal, external and in use aspects of each external quality characteristic. It enumerates the main functionalities of Quality Monitoring Tool that automates the assessment activities. This is followed by the conclusion in Section 4.

## 2. Quality characterization and assessment

### 2.1 External quality of inter organizational systems

Organizations are increasingly concerned with the quality management of their information systems and make continuous investments to enhance their different qualities levels. These investments are justified by the importance of ISQ in order to achieve organizational objectives and to increase performance and profits [1].

Also, there is an increasing trend for several information systems to span boundaries between organizations. Such systems can be used to support collaborations and partnerships among organizations for competitive purposes. Low quality level of inter organizational systems is a potential failure of cooperation and collaboration [2].

Within a collaborative ecosystem, ISQ improvement deals with conceptual, organizational and technical barriers between stakeholders that may belong to different governance subdomains [3].

For this purpose, Studying ISQ implies a specific focus on its external characteristics that are influenced by environmental parameters and impacted by the ecosystem factors. As example of such characteristics, we quote (i) interoperability, (ii) security, (iii) adaptability, (iv) flexibility, (v) horizontal alignment ability.

Many efforts are done in order to propose approaches to ensure, manage and control ISQ characteristics. In this area, one of the active branches deals with the characterization and the assessment techniques [4]. Indeed, many models are proposed to characterize quality attributes and their relationships. Despite the richness and benefits of these models, they present several disadvantages regarding the consideration of all operational aspects in assessment [3, 5].

## 2.2 Quality models

The domain of information system quality is subject to numerous modeling initiatives. The first ones lead to hierarchical definitions of quality factors composed of characteristics, which lead to evaluations based on metrics [6, 7, 8]. Such models represent specifications of ISQ characteristics. They (i) relate various quality attributes, (ii) identify practices to address them and (iii) describe metrics for measuring or observing them [9].

ISQ models are useful in (a) the requirements identification and their completeness validation processes, (b) the identification of design and testing objectives and the user acceptance criteria and (c) the communication improvement between all information system stakeholders (acquirers, architects, developers, etc.) [9].

Information system engineering researchers and practitioners have suggested many different quality models. The well-adopted ones characterize ISQ as a hierarchical multidimensional system [10]. There are also other models that adopt relational topology [11].

Other models exist and are star-based topology [12] or have Bayesian Belief Network (BBN) topology [13]. In spite of the number of similarities, each proposed model has its own terminology and has a varying number of attributes as illustrated in Table 1 below.

ISO/IEC 9126-1[14] quality model incorporates the findings of previous models. ISO/IEC 9126-1 includes six characteristics (Functionality, Reliability, Usability, Efficiency, Maintainability, And Portability), which are further subdivided into 27 sub-characteristics. For instance, functionality characteristic includes a set of sub characteristics: security, interoperability, suitability, accuracy, compliance with standards.

Table. 1: ISQ models components

| ISQ Model | Topology | First / Second Level | First level Characteristics |
|---|---|---|---|
| Mccall (1976) [6] | H* | Factor / Criteria | Correctness, reliability, efficiency, integrity, usability, maintainability, testability, flexibility, portability, reusability, and interoperability |
| Boehm (1978) [7] | H* | High level charac.*/ Primitive charac. | Testability, understandability, efficiency, Modifiability, reliability, portability, and human Engineering |
| Murine (1983)[15] | H* | Factor / Criteria | Correctness, Reliability, Efficiency, Integrity, Reusability, Usability, Maintainability, Testability, Flexibility, Portability, Interoperability, Intraoperability |
| Bowen (1985)[16] | H* | Factor / Criteria | correctness, Reliability, Efficiency, Usability, Integrity, Maintainability, verifiability, Portability, flexibility, reusability, interoperability, survivability, expandability |
| Evans and Marciniak (1987)[17] | H* | Factor / Criteria | Mccall (1976) [6] factors + Verifiability, Expandibility |
| FURPS (1987)[18] | H* | Charac./ Sub charac. | Functionality, usability, reliability, performance, supportability. |
| Gillies (1987)[19] | R* | - | Maintainability, Flexibility, Testability, Portability, Reusability, Interoperability, Correctness, Reliability, Efficiency, Integrity, Usability |
| Deutsch and Willis (1988)[20] | R* | Factor / Criteria | Clarity, Integrity, Traceability, Reliability |
| ISO/IEC 9126 (1991)[14] | H* | charac. / Sub charac. | Reliability, maintainability, Portability, usability, functionality, and efficiency |
| Dromey (1996)[21] | H* | H-level attribute / Sub attribute | Maintainability, Reliability, efficiency, usability, portability, Reusability, functionality |
| IEEE 1061 (1998)[22] | H* | Factor / Subfactor | Efficiency, functionality, maintainability, portability, reliability, usability |
| Perry (1987)[11] | R* | - | Correctness, Reliability, Efficiency, Integrity, Usability, Maintainability, Testability, Flexibility, Portability, Reusability, Interoperability |
| Stefani et al. (2003)[13] | B* | - | Operability, Reliability, Functionality, Efficiency |
| Khosravi (2004)[12] | S* | - | Usability, Understandability, Learnability., Operability, Flexibility, Reusability, Robustness, Environmental tolerance, error tolerance, failure tolerance, scalability |
| AOSQUAMO (2009)[23] | H* | charac. / Sub charac. | ISO/IEC 9126 charac. |

H* : Hierarchical, R* : Relational, S*: Star based, B*: BBN based
charac.*: characteristic

In fact, ISQ can be viewed from various perspectives. Several taxonomies have been proposed in this context. In this sense, there are (see Fig 1):
- Many levels of ISQ concern: business, process, service and data level [24].
- Various approaches to establish ISQ: integrated, federated, and unified approach [25].
- Multiple barriers could handicap ISQ establishment: conceptual, organizational and technical barriers [26].
- Different scopes of application: within the same organization, cross independent organizations [27],
- Different ISQ aspects: internal, external and in use aspects [28].

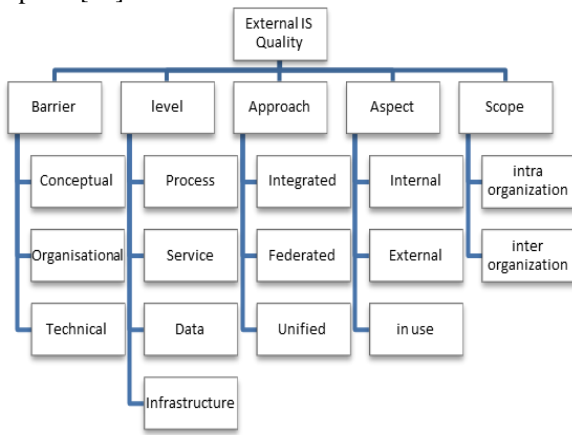

Fig. 1. EISQ taxonomy

2.3 Assessment approaches

In addition to characterization concern, one of the active branches in ISQ management discipline is assessment methods and metrics development. ISQ models, and especially those who belong to "Factor, Criteria, Metric" Family, note the importance of metrics in order to assess quality levels. Also, ISQ management discipline denotes the development of several measurement approaches for the same characteristic. As an illustration, Elmir et al. [29] identifies fourteen different assessment approaches for the single characteristic of interoperability. These approaches have many differentiation aspects like (i) domain of application, (ii) system orientation (qualifying a system or a link between systems), (iii) evaluation instance (a priori, a posteriori), (iv) level character, (v) quantitative or qualitative evaluation, and (vi) coupling with advanced mathematic techniques [29].

In spite of the richness and the advantages of existing assessment approaches, they present several disadvantages regarding the consideration of all operational aspects in assessment. Therefore:

1. The majority of approaches assess "a priori" ISQ degree. Few of them are interested on "a posteriori" aspects. No one of them proposes to take into consideration the both aspects.
2. The existing approaches qualify quality degree of a component or a link between systems. Few of them are able to take in charge the two situations.
3. The majority of approaches is qualitative and describes maturity level with a specific perspective (technical or organizational).
4. Few of the existing approaches combine their results with advanced mathematical techniques such optimization, probability, matrixes (linear algebra), logics or complexity.
5. None of the approaches explicit how prior assessment is used for effective implementation of planned state. Indeed, the measurement process stops when the ISQ degree is calculated. Using this level thereafter as explicit parameter improvement is supported by no one of the existing approaches.

## 3. RatQual Model: an assessment-oriented Quality model

RatQual (for Ratio of Quality) is an assessment oriented model that proposes an innovative three axis hierarchical classification of 17 EISQ characteristics. The result of the assessment approach is a ratio metric enabling the measurement of specific EISQ characteristic by taking into account three main operational aspects: internal, external and in use ones.

### 3.1 EISQ RatQual categorization

This work proposes a classification of EISQ characteristics using a requirements engineering perspective. Quality requirements engineering is a discipline interested into formal quality requirements definition and change management.

Requirements engineering applied to quality management area and specifically to EISQ extent implies a three axis categorization. The first axe is about the functional ISQ identification. The two other axes are related to change requests issues. Indeed, the second axe is more interested into context dependent adaptation requests. The third axe is more sensitive to requests evolution over time.

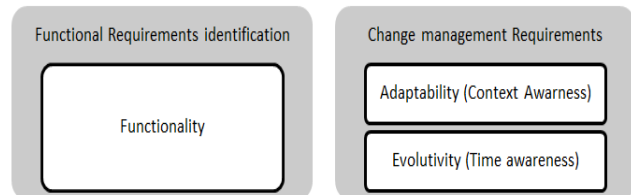

Fig. 2. Requirement perspective in EISQ Classification

Indeed, the first class of characteristics is Functionality. This class refers to the essential purpose of the involved information systems and their components. Functionality characteristics are mainly recognized in the requirements identification stage. This class contains various features among which interoperability, security, compliance and inter-alignment ability.

The other classes we propose are related to quality requirements linked to system change management. Change requests can be classified into two main categories: (i) "Adaptability category" including context dependent change requests, and (ii) "evolutivity category" time dependent change requests.

The former EISQ category entitled "Adaptability" includes Portability, Coexistence, Replace ability, Flexibility and Variability. The latter category named "Evolutivity" encloses characteristics like Changeability, Maintainability, Stability, Testability, Customizing Ability and Extensibility.

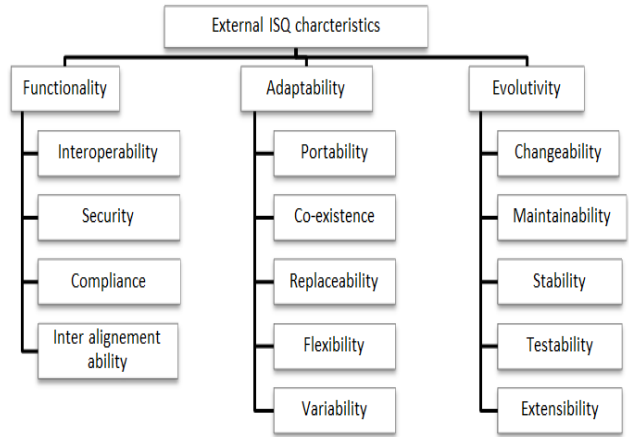

Fig. 3. EISQ characteristics classification

### 3.2 EISQ appraisal aspects

In terms of appraisal dimensions, ISQ characteristics have a priori and a posteriori assessment details. A posteriori assessment is related to in use operational performance appraised after effort implementation of the desired quality. A priori assessment, for its part, takes into consideration two features: (i) quality potentiality and (ii) quality implementation compatibility.

For this purpose, the proposed model identifies three essential appraisal aspects:

• Quality potentiality: it is an «internal feature» of a system that reflects its characteristic preparation within a collaborative context. This involves identifying a set of requirements that have an impact on interaction characteristic capacity with partner's systems without necessarily having concrete information on them. The objective is to foster quality readiness and preparation by eliminating barriers that may reduce the quality degree.

•Quality implementation compatibility: it represents an «external feature». In fact, enhancing the characteristic ability of two support systems is ensured through an engineering process aiming to establish inter organizational collaboration between them and also respond to the desired characteristic requirements.

•Quality performance: the third aspect characterizes the «quality in use». It focuses on monitoring operational performance. It consists of an assessment of the communication infrastructure availability, and the supporting system in general.

### 3.3 RatQual Model

RatQual is an assessment oriented model intended to describe external characteristics that are influenced by environmental. RatQual aims to evaluate EISQ quality characteristics using a ratio metric.

RatQual metric aggregates a set of sub metrics that asses complementary aspects. These aspects include "a priori" and "a posterior" aspects. A priori aspects consist of internal aspects in one hand and external aspects on another hand.

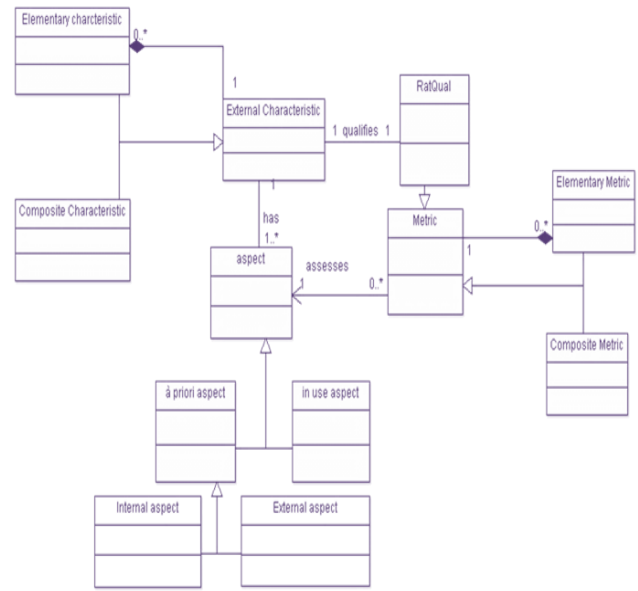

Fig.4. RatQual characterization and assessment model

# 4. RatQual Assessment Approach

RatQual is a five steps appraisal approach. These steps are as follows (see Figure 5):
1. Delineating the scope of the study.
2. Quantifying the internal aspect: quality characteristic potentiality.
3. Calculating the external aspect: Quality implementation effort.
4. Evaluating the in use aspect: operational quality performance.
5. Aggregating the EISQ RatQual degree based on an adequate aggregation technique.

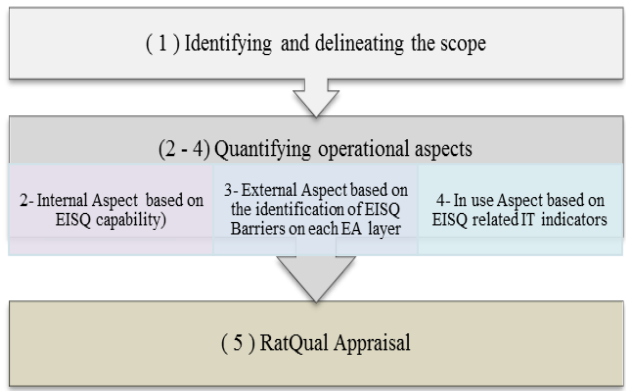

Fig.5. RatQual assessment approach

## 4.1 Scope delineation

Assessing an external quality characteristic degree of a system requires the knowledge of its ecosystem. In practical terms, the study focuses on a macro business process consisting of a set of sub auto-mated processes among independent business entities. These sub processes are linked together via several interfaces identified in advance. In this case, the preliminary phase consists of identifying the context of the studied automated business process then lists its underlying automated processes. This step includes identifying:
- Organizations involved in the cooperation.
- Sub process within each entity in order to study the compatibility degree of stakeholders in order to enhance a specific EISQ.
- Information systems that support automated business processes within each organization.
- Application services that enables sub processes collaboration.

## 4.2 Internal aspect: Quality potentiality

The calculation of the potential for quality characteristic within the $k_{th}$ organization «$QP_k$» requires the adoption of one of the quality maturity models. The organization is classified then on one of the five levels noted QMML (for Quality maturity model level). Table 2 below illustrates a EISQM inventory.

Table. 2. Information System Quality Maturity Models

| Quality charac. | Quality maturity model |
|---|---|
| Functionality | Functionality maturity model integration (FMMI) [30,31] |
| Interoperability | Interoperability Maturity Model (IMM), Enterprise IMM (EIMM), Organizational IMM (OIMM), Level of Information System Interoperability (LISI) [3, 5, 29] |
| Security | Information Security Maturity Model (ISMM)[32] |
| Compliance | Governance Compliance Maturity Model (GoCoMM)[33] |
| Inter alignment ability | Inter alignment ability maturity model (IAMM) [34] |
| Adaptability | Quality maturity model (QMM) [35], Adaptability maturity model Integration (AMMI) [30] |
| Portability | Portability maturity model Integrtaion (PMMI) [30] |
| Co-existence | QMM [35] |
| Replace ability | QMM[35] |
| Flexibility | Flexibility maturity model (FMM)[36] |
| Variability | QMM [35] |
| Evolutivity | QMM [35] |
| Changeability | QMM [35] |
| Maintainability | Architecture Maintainability Maturity Model (AM3) [37], QMM [35] |
| Stability | QMM [35] |
| Testability | Testability maturity model (TMM)[38] |
| Extensibility | QMM [35] |

To identify the potential degree of a specific quality characteristic, we propose then the following mapping (See Table 3):

Table. 3. Quantification of quality maturity

| Maturity Level (QMML) | Potentiality quantification |
|---|---|
| 1 | 0.2 |
| 2 | 0.4 |
| 3 | 0.6 |
| 4 | 0.8 |
| 5 | 1 |

Within each organization, the potential is calculated using the following equation

$$QP_k = 0.2 * QMML_k \qquad (1)$$

The final characteristic potentiality is given by Equation 2 below:

$$QP = \min(QP_k) \quad (2)$$

### 4.3 External aspect: Quality compatibility

To assess the external aspect degree, the present work uses a compatibility matrix [3, 5, 29].
The compatibility matrix, as presented in Table 3, consists of a combination of the "quality levels perspective" and "quality barriers perspective" depicted in Fig.1 of section 2.B. In practical terms, we enumerate conceptual, technical and organizational barriers in the different layers of collaboration concern: process, service, data and infrastructure.
By noting the elementary degree of interoperation compatibility «$dc_{ij}$» (i takes values from 1..4, and j takes values from 1..6).

Table. 3. Interoperation compatibility

|  | Conceptual | | Organizational | | Technology | |
|---|---|---|---|---|---|---|
|  | Syntactic | Semantic | Responsibilities | Organization | Platform | Communication |
| Process | $dc_{11}$ | $dc_{12}$ | $dc_{13}$ | $dc_{14}$ | $dc_{15}$ | $dc_{16}$ |
| Service | $dc_{21}$ | $dc_{22}$ | $dc_{23}$ | $dc_{24}$ | $dc_{25}$ | $dc_{26}$ |
| Data | $dc_{31}$ | $dc_{32}$ | $dc_{33}$ | $dc_{34}$ | $dc_{35}$ | $dc_{36}$ |
| Infrastructure | $dc_{41}$ | $dc_{42}$ | $dc_{43}$ | $dc_{44}$ | $dc_{45}$ | $dc_{46}$ |

Therefore, if the criteria in an area marked satisfaction the value 0 is assigned to dcij; otherwise if a lot of incompatibilities are met, the value 1 is assigned to dcij.
The degree of compatibility «DC» is given as follows:

$$DC = 1 - \sum (dc_{ij}/24) \quad (3)$$

### 4.4 In use aspect: operating performance

By Denoting:
«DS» the overall availability rate of application servers.
«QoS» service quality of different networks used for interacting components communication. QoS is represented mainly by the overall availability of networks.
«TS» end users satisfaction level about interoperation.
Given the cumulative nature of these three rates, the evaluation of operational performance is given by the geometric mean [31] as the following equation (See Equation 4):

$$PO = \sqrt[3]{(DS * QoS * TS)} \quad (4)$$

### 4.5 RatQual aggregation

The final calculation of RatQual (for ratio of Quality) is by aggregating the three previous indicators using a function f defined in $[0,1]^3 \rightarrow [0,1]$ (See Equation 5)

$$RatQual = f(PQ, DC, PO) \quad (5)$$

Given the independent nature of these three indicators, we opt for the arithmetic mean [31] as follows (See Equation 6):

$$RatQual = (PQ + DC + PO)/3 \quad (6)$$

In case we have elements for pondering each one of these three indicators with different weights (w1, w2, w3); we choose the weighted arithmetic mean.

$$RatQual = (w1*PQ + w2*DC + w3*PO)/(w1+w2+w3) \quad (6)$$

### 4.6 Quality monitoring Tool (QMT)

The Quality monitoring tool (QMT) automates the RatQual assessment approach. It includes three principal modules. The first one is dedicated to EISQ characteristic assessment at a specific period. The second one proposes a viable scheme to reach a planned Quality degree. The third module includes a set of reporting views designed to enable periodical quality monitoring activities.

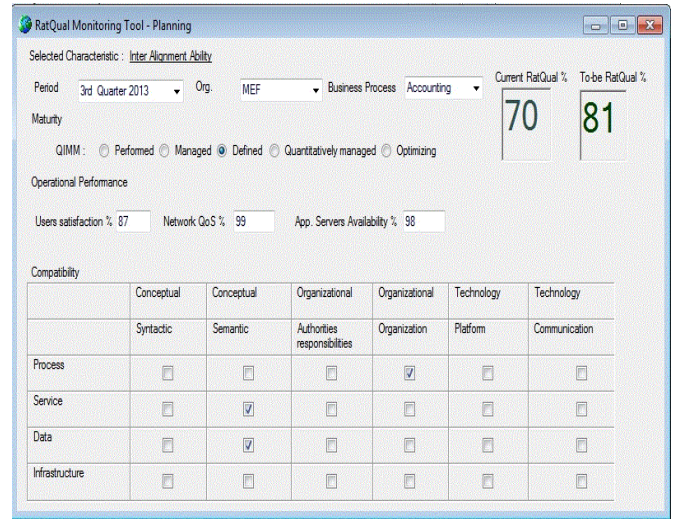

**Fig.5. Quality monitoring tool (QMT)**

Indeed, QMT has the capacity to track periodically the evolution of quality degree. It gives the possibility to propose a scenario to reach a planned degree of quality characteristic. For instance, in the example shown on Fig. 5, we plan to increase the "inter alignment ability" ratio from an "As-is" degree to a "To-be" one. QMT proposes to (i) improve horizontal alignment maturity to reach the third stage, (ii) optimize the availability of involved application servers, (iii) better meet end users expectations and (iv) to resolve semantic incompatibilities.

## 4. Conclusions

To operate effectively, organizations are encouraged to enter into close interaction with all their partners. Inter-organizational collaboration is a strategic issue. Quality assurance in this context is very important. For this purpose, this work proposes a novel assessment oriented model for context-dependent quality. This model, named RatQual, takes into account conceptual, organizational and technical considerations and gives importance to architectural elements.

The proposed model here serves to characterize information system external qualities that are influenced by environmental parameters. RatQual considers internal, external and in use aspects. It combines à priori evaluation elements within the design phase of interconnection setup and à posteriori evaluation aspects considering the performance degree of collaboration.

**A. ELMIR** "Ph.D. candidate" at ENSIAS (National Higher School for Computer Science and System analysis) Rabat, holder of a master degree from ENIM Rabat and a mastere diploma from INSA Lyon France (2011). Her research focuses on multi objective optimization and optimal control of information system quality within collaborative networks. She is an integration architect on a private Financial Holding (Banking, Insurance). She also runs the Solutions support activity of this holding.

**B. ELMIR** He received a Ph.D. degree (2012), an Extended Higher Studies Diploma (2006) and a Software engineer degree (2002) from ENSIAS, (National Higher School for Computer Science and System analysis), Rabat. His research currently focuses on interoperability optimization on public administration. He is an integration architect on the Ministry of Economy and Finance of Morocco since 2002. He also oversees information system quality assurance activity in this department.

**B. BOUNABAT** Ph.D. in Computer Sciences. Professor in ENSIAS, (National Higher School for Computer Science and System analysis), Rabat, Morocco. International Expert in ICT Strategies and E-Government to several international organizations, Member of the board of Internet Society - Moroccan Chapter.